\documentclass[aps,twocolumn,superscriptaddress]{revtex4-1}
\usepackage[colorlinks=true,bookmarks=true,citecolor=magenta,urlcolor=magenta,linkcolor=magenta,breaklinks]{hyperref} 
\usepackage{breakurl}
\usepackage{sidecap}
\usepackage{amssymb}
\usepackage{hhline}
\usepackage{urwchancal}
\sidecaptionvpos{figure}{t}
\usepackage{amsmath}
\usepackage{graphicx}
\usepackage{esint}
\usepackage{epstopdf}
\usepackage{rotating}
\graphicspath{{pict/}{./}}
\usepackage{bm}%
\usepackage{txfonts}
\usepackage[classicReIm]{kpfonts}

\newcounter{Fig}

\begin{document}


\title{Global Mie Scattering}

\author{Weijin Chen}
\email{Authors contributed equally to this work.}
\affiliation{School of Optical and Electronic Information, Huazhong University of Science and Technology, Wuhan, Hubei 430074, P. R. China}
\author{Qingdong Yang}
\email{Authors contributed equally to this work.}
\affiliation{School of Optical and Electronic Information, Huazhong University of Science and Technology, Wuhan, Hubei 430074, P. R. China}
\author{Yuntian Chen}
\email{yuntian@hust.edu.cn}
\affiliation{School of Optical and Electronic Information, Huazhong University of Science and Technology, Wuhan, Hubei 430074, P. R. China}
\author{Wei Liu}
\email{wei.liu.pku@gmail.com}
\affiliation{College for Advanced Interdisciplinary Studies, National University of Defense
Technology, Changsha, Hunan 410073, P. R. China}

\begin{abstract}
In various subdisciplines of optics and photonics, Mie theory has been serving as a fundamental language and play indispensable roles widely. Conventional studies related to Mie scattering largely focus on local properties such as differential cross sections and angular polarization distributions. Though  spatially integrated  features of total cross sections in terms of both scattering and absorption are routine for investigations, they are intrinsically dependent on the specific morphologies of both the scattering bodies and the incident waves, consequently manifesting no sign of global invariance. Here we propose global Mie scattering theory to explore topological invariants for characterizations of scatterings by any obstacles of arbitrarily structured or polarized coherent light. It is revealed that, independent of distributions and interactions among the scattering bodies of arbitrary geometric and optical parameters, in the far field inevitably there are directions where the scatterings are either zero or circularly polarized.  Furthermore, for each such singular direction we can assign a half-integer index and the index sum of all those directions are bounded to be a global topological invariant of $2$.  The global Mie theory we propose, which is mathematically simple but conceptually penetrating, can render new perspectives for light scattering and topological photonics in both linear and nonlinear regimes, and would potentially shed new light on the scattering of acoustic and matter waves of various forms.
\end{abstract}

\maketitle

\section{Introduction}
The seminal problem of light scattering by particles and the associated Mie theory has been pervasive in every subject of photonics, laying the foundation for not only researches and applications in optics and physics~\cite{Bohren1983_book,DOICU_light_2006,GOUESBET_generalized_2011}, but also for those in many other interdisciplinary fields including even biology and medicine~\cite{FERRARI_2005_NatRevCancer_Cancer,DREADEN_2012_Chem.Soc.Rev._golden}. Conventional studies based on Mie theory concentrate on properties that can be roughly divided into two categories: overall integrated ones such as total absorption, scattering and extinction cross sections; local ones such as differential scattering cross sections and angular polarization distributions. Mie theory and those scattering properties broadly underlie various topics of nanophotonics, especially the new concepts and phenomena of cloaking and invisibility~\cite{Alu2009_PRL}, superscattering~\cite{Ruan2010_PRL,QIAN_2019_Phys.Rev.Lett._Experimental}, Kerker scattering and its generalized forms~\cite{Kerker1983_JOSA,LIU_2018_Opt.Express_Generalized}, giving birth to and further expanding the field of Mie-Tronics that largely originate from interferences of radiating multipoles of different orders and natures~\cite{jahani_alldielectric_2016,KUZNETSOV_Science_optically_2016,WON_2019__Mietronic}.

Besides the rapid progress relying on Mie theory in various directions, photonics at the same time has gained great momentum through incorporating novel topological concepts ~\cite{Lu2014_topological,OZAWA_2018_ArXiv180204173}. New topology-related ideas from condensed matter physics and other branches of physics have rendered  extra degree of freedom for manipulations of light matter interactions, through comprehensive exploitations of topological properties that are globally bounded~\cite{Lu2014_topological,OZAWA_2018_ArXiv180204173}. For the classical scenario of light scattering by arbitrary obstacles, at the first glance, the identification of globally invariant properties seems to be out of reach. This is simply due to the fact that both the aforementioned overall and the local scattering properties are intrinsically dependent on geometric and optical parameters of the specific scattering bodies, their distribution patterns and the cross interactions~\cite{Bohren1983_book,DOICU_light_2006}. Moreover, those scattering features are also dependent on the morphologies of the incident waves, especially their spatial and polarization contextures~\cite{GOUESBET_generalized_2011}. Apart from those seemingly insurmountable difficulties, recently both local and globally invariant topological features have been revealed for radiating electromagnetic multipoles of arbitrary orders, revealing a hidden dimension of Bloch modes in periodic structures and their topological charges~\cite{CHEN_2019__Singularities,CHEN_2019_ArXiv190409910Math-PhPhysicsphysics_Linea}. As essential entities in Mie theory,  it is well known that electromagnetic multipoles serve as a orthogonal and complete basis for radiation expansion of any sources. As a result, it is natural to expect that the same topological idea can be extended to Mie scattering configurations with arbitrary finite obstacles and any coherently polarized incident waves.

In this work we propose \textit{Global Mie Scattering theory} to explore topological properties that are globally invariant for arbitrary Mie scattering configurations. Regardless of the specific morphologies of both the incident waves and the obstacles, the scatterings can always be mapped on the momentum sphere as tangent (transverse) fields perpendicular to the scattering directions. For such continuous tangent fields on a two dimensional spherical surface, the fundamental Poincar\'{e}-Hopf theorem can be directly applied~\cite{MILNOR_1997__Topology,NEEDHAM_1998__Visual}. As for our physical problem of Mie scattering of any specific configurations, this mathematical theorem secures the following two facts: (i) Across the momentum sphere, despite some obviously trivial cases such as invisibility of zero total scattering, there must be isolated directions along which the scattering is either zero or circularly polarized; (ii) For each of such singular directions, a half-integer index can be assigned and the index sum of all those directions are bounded to be $2$, which is the Euler characteristic of the momentum sphere.  Our mathematically simple while conceptually deep Global Mie theory has effectively blended two sweeping concepts of Mie scattering and  global topology of Euler characteristics and singularities~\cite{MILNOR_1997__Topology,NEEDHAM_1998__Visual}. Considering the foundational roles of Mie theory and the recent rapid pervasion of topological concepts throughout photonics,  we believe that the proposed global theory can further accelerate the interplay of photonics and topology. This can hopefully make possible more flexible controls of light-matter interactions in both linear and nonlinear regimes, with broad implications for molecular scattering and optical activity studies, and even for waves of other forms.

\section{Linearly Polarized Scattering by Individual Particles}
\begin{figure}[tp]

\centerline{\includegraphics[width=9cm]{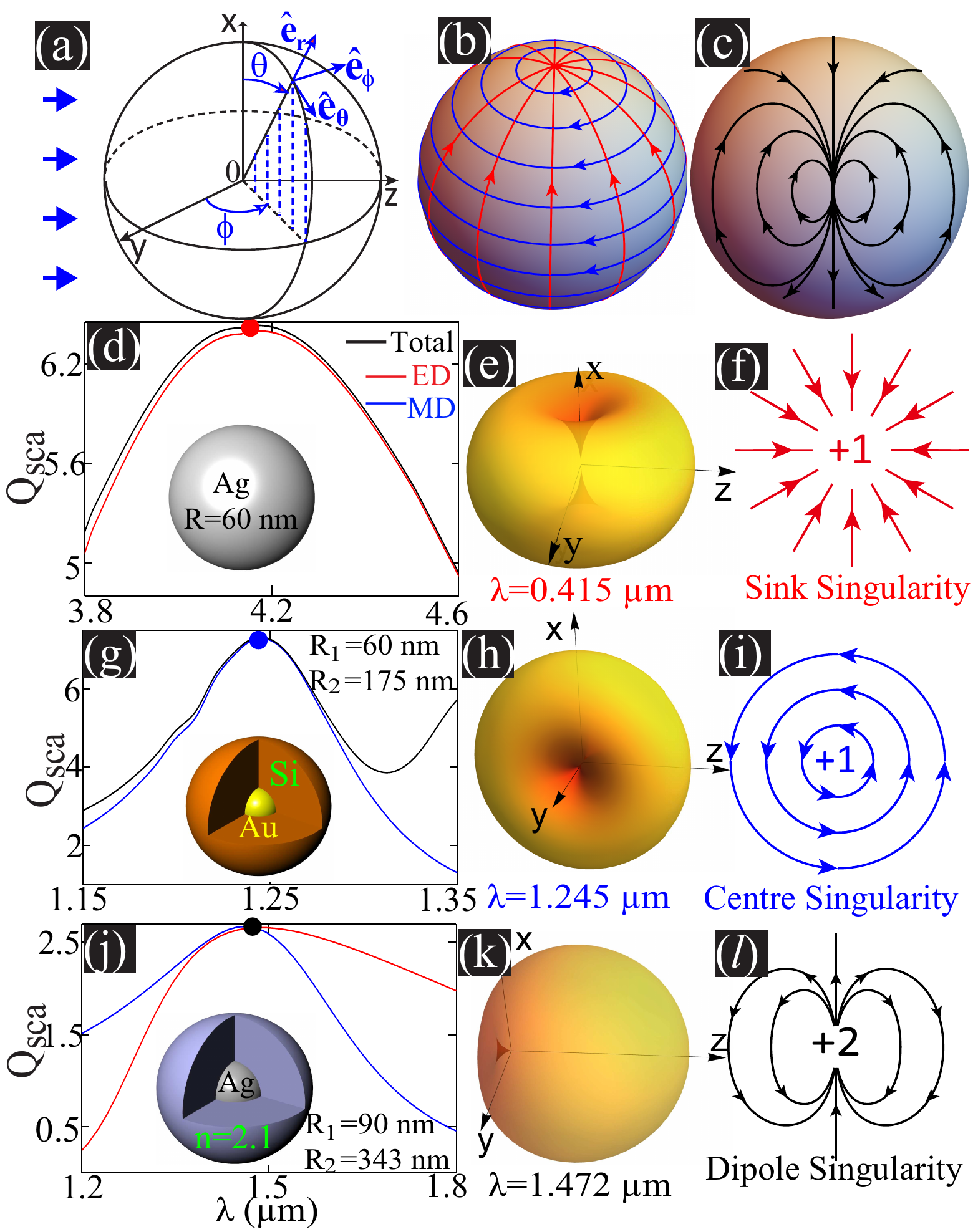}} \caption {\small (a) The momentum sphere parameterized by both Cartesian and polar coordinates,  with the associated orthonormal basis vectors $\mathbf{\hat{e}}_\theta$, $\mathbf{\hat{e}}_\phi$ and $\mathbf{\hat{e}}_r$ specified. Other parameters have their usual meaning. On the momentum sphere, continuous tangent vectors exist with two singularities of Poincar\'{e} index $+1$ (located on poles) in (b), and one singularity of Poincar\'{e} index $+2$ in (c). (d), (g) and (j): Scattering efficiency spectra (both total and dipolar partial ones are included) for homogeneous or core-shell spherical particles (see insets) with the resonant positions pinpointed by dots. The scattering patterns at those positions are shown correspondingly in (e), (h) and (k). Vectorial field patterns close to singularities are shown in (f), (i) and ($l$), the Poincar\'{e} indices of which are also specified.}\label{fig1}
\end{figure}

For the general case of light scattering by arbitrary finite obstacles, regardless of the shapes and optical properties of each individual scattering body, and how different scattering bodies are clustered, the scattered fields are continuously distributed among different directions on a momentum sphere parameterized by $\theta$ and $\phi$ [see Fig.~\ref{fig1}(a) for the momentum sphere in a spherical coordinate system].  Moreover, since in the far field the scattering is purely transverse with both electric and magnetic fields perpendicular to the scattering directions (denoted by $\mathbf{\hat{e}}_r$), the fields can be viewed as continuous tangent fields on the $\mathbf{\hat{e}}_\theta$-$\mathbf{\hat{e}}_\phi$  plane.

A simple scenario to start with is that all the scattered fields are linearly polarized throughout the mentum sphere, on which the tangent electric and magnetic fields can be fully represented by vector fields. As for such continuous vector field on the momentum sphere, the Poincar\'{e}-Hopf theorem~\cite{MILNOR_1997__Topology,NEEDHAM_1998__Visual} requires that there must be isolated directions where there is no scattering (the vector field and also the total intensity are strictly zero). Those directions correspond to singularities of vector fields (vectorial singularities denoted by \textbf{V} points in this work), and for each singularity an integer Poincar\'{e} index (or equivalently winding numbers of the vector fields that trace out the closed contour~\cite{NEEDHAM_1998__Visual}) can be assigned.  All across the momentum sphere, the  sum of Poincar\'{e} indices  for all singularities are bounded to be $2$, which is the corresponding Euler characteristic~\cite{MILNOR_1997__Topology,NEEDHAM_1998__Visual}. In Fig.~\ref{fig1}(b) we show two sets of well known vector fields consisting of latitude ($\theta$ is constant) and longitude ($\phi$ is constant) vectors. For both sets, there are two singularities locating on the poles ($\theta=0$ or $\pi$), each of Poincar\'{e} index $+1$ with their sum being $2$, as required by the Poincar\'{e}-Hopf theorem. We have also shown in Fig.~\ref{fig1}(c) another set of less well known but equally important vector fields with only one singularity of Poincar\'{e} index $+2$, which is termed as a dipole singularity (the  vectorial patterns close to the singularity are the same as those of a static electric dipole)~\cite{NEEDHAM_1998__Visual}.

The simplest scenario of everywhere linearly polarized scattering is that by an obstacle which can be viewed as a linearly oscillating electric dipole (ED). Rayleigh scattering with linearly polarized incident plane waves falls into this category~\cite{Bohren1983_book}.  In  Fig.~\ref{fig1}(d), alternatively, we show another example of a plane wave scattering by an Ag sphere (radius $R=60$~nm; permittivity adopted from Ref.~\cite{Johnson1972_PRB}) that supports surface plasmon resonances~\cite{Maier2007}. The plane wave is $\textbf{x}$-polarized and propagating along $\textbf{z}$, and the scattering spectra (scattering efficiency $Q_{\rm sca}$, that is scattering cross section divided by the cross section of the sphere) in terms of both total and partial ED scatterings are summarized in Fig.~\ref{fig1}(d). The geometric and optical properties of the scattering bodies can be found in the figure, as is the case for other scattering configurations throughout this work. As is shown, at the marked resonant position (the wavelength $\lambda=415$~nm), there is ED scattering which is linearly polarized throughout the momentum sphere. The far-field scattering pattern  (in terms of angular scattering intensity, as is the case throughout this work) at this spectral position is shown in Fig.~\ref{fig1}(e), with their singularities locating at the poles, which coincide with the dipole oscillating direction. A simple analysis based on the formulas of dipolar scattered fields reveals that scattered magnetic and electric fields of the ED on the momentum sphere are exactly represented by those latitude and longitude vectors shown in Fig.~\ref{fig1}(b), respectively~\cite{jackson1962classical}. The electric field patterns at the neighbourhood of the singularities are shown in Fig.~\ref{fig1}(f). Here only the sink singularity of Poincar\'{e} index $+1$ at one pole is shown and the corresponding source singularity at the opposite pole can be directly obtained through flipping the orientations of all vectors, with the same index $+1$~\cite{NEEDHAM_1998__Visual}.

The duality of Maxwell Equations guarantees that the scattering pattern of an ED and magnetic dipole (MD) would be identical, with the interchange of electric and magnetic terms (including both the dipolar moments and vector fields)~\cite{jackson1962classical}. Figure \ref{fig1}(g) shows the scattering spectra of an Au-Si core-shell particle, where for the gold core $R_1=60$~nm and permittivity taken from Ref.~\cite{Johnson1972_PRB}, and for the silicon shell $R_2=175$~nm and permittivity taken from Ref.~\cite{PALIK_1998__Handbook}. At the resonant position ($\lambda=1245$~nm) a pure MD can be exclusively obtained~\cite{liu_toroidal_2015,FENG_2017_Phys.Rev.Lett._Ideal}.  Apart from a $\pi/2$ rotation about $\textbf{z}$ axis (since the MD moment is along $\textbf{y}$ axis parallel to the incident magnetic field, while the ED moment is along $\textbf{x}$ axis parallel to the incident electric field), the scattering pattern [shown in Fig.~\ref{fig1}(h)] is the same as that of an ED and electric field patterns close to the singularities (centre singularities located at $\phi$=0 and $\pi$ with Poincar\'{e} index $+1$) are shown in Fig.~\ref{fig1}(i). According to the duality principle, the patterns of tangent electric and magnetic fields of the MD are exactly the same as those of tangent magnetic and electric fields of the ED, respectively.  They are respectively latitude and longitude vectors shown in Fig.~\ref{fig1}(b), meaning that the magnetic field singularities of MD are source or sink singularities.

Another elementary case of linearly polarized scattering is the one that can be represented by vector fields shown in Fig.~\ref{fig1}(c), with only one dipole singularity of Poincar\'{e} index $+2$. This type of field patterns are inaccessible to obstacles that support a single multipolar moment, the scattering of which is symmetric with at least two singularities. The simplest configuration to achieve this is based on an in-phase overlapping of an ED and a MD of the same magnitude~\cite{LIU_2018_Opt.Express_Generalized,Liu2012_ACSNANO,CHEN_2019__Singularities}, and this ED-MD pair is also termed as a dynamic Kerker dipole~\cite{Kerker1983_JOSA,LIU_2018_Opt.Express_Generalized,CHEN_2019__Singularities}. Figure \ref{fig1}(j) shows the scattering spectra of a core-shell particle [the core is Ag of radius $R_1=90$~nm and the shell is dielectric (refractive index $2.1$) of radius $R_2=343$~nm], which confirm that at the marked resonant position ($\lambda=1472$~nm) the ED and MD overlap and the particle can be treated as a Kerker dipole.   The scattering  pattern at the overlapping resonant position [see Fig.~\ref{fig1}(k)] exhibits a dipole singularity at the minus $\textbf{z}$ backward direction, with the vector field  pattern close to it shown in Fig.~\ref{fig1}($l$) which is identical to that shown in Fig.~\ref{fig1}(c).

\section{Arbitrarily Polarized Scattering by Individual Particles}

Up to now, we have discussed only the simple scenario when the scattering is fully linearly polarized all across the momentum sphere. Nevertheless, it is well known that light is generally elliptically polarized, and the linear and circular polarizations are merely special cases~\cite{jackson1962classical}.  For general scattering of elliptic polarizations, at any fixed temporal moment we can treat the scattered fields as vector fields [as presented already in Fig.~\ref{fig1}], and then the conclusions drawn above for vectorial singularities of integer Poincar\'{e} indices can be directly applied.  However, the vectors are rotating temporally for elliptic polarizations, meaning that the vector field patterns are constantly changing. To effectively characterize the evolving patterns and to further establish connections to experimentally measurable quantities that are time-averaged, static line fields should be introduced~\cite{CHEN_2019_ArXiv190409910Math-PhPhysicsphysics_Linea,HOPF_2003__Differential,NYE_natural_1999,GBUR_2016__Singular}. Intuitively, line fields are non-oriented vector fields, that is, lines without arrows~\cite{HOPF_2003__Differential,NYE_natural_1999,GBUR_2016__Singular}. Transversing a closed contour within vector fields would bring a vector back to itself with a fixed orientation, requiring that the overall rotation angle along the loop being an integer number of $2\pi$ and thus leading to integer Poincar\'{e} indices; While for line fields, since there is no orientation for each consisting line, a half-integer number of $2\pi$ rotation is sufficient to bring the line field back to itself, resulting in half-integer Hopf indices~\cite{HOPF_2003__Differential}.  The intrinsic singularities of vector fields are those of Poincar\'{e} indices $\pm1$, and other higher-index ones can be viewed as a singularity composite when multiple intrinsic singularities overlap with one another. The same principle is also applicable to line fields, but here the intrinsic singularities are those of Hopf indices $\pm1/2$. From this perspective, for both line and vector fields, the higher-index singularities are accidental, which can be simply decomposed into several intrinsic singularities through further perturbations~\cite{MILNOR_1997__Topology,NEEDHAM_1998__Visual,HOPF_2003__Differential}.

For electromagnetic waves, the widely constructed line fields for general elliptical polarizations consist of long axes of the polarization ellipses, which are also termed as polarization fields~\cite{NYE_natural_1999,GBUR_2016__Singular}.  For linear polarizations, line fields can be directly constructed through removing the arrows of the vectors. Then the vectorial singularities become line singularities automatically, with the Poincar\'{e} and Hopf indices equal to each other.  For example, the vector fields (vectorial singularities) shown in Figs.~\ref{fig1}(b) and (c) would become line fields (line singularities) with the arrows removed, with the positions and indices of the singularities unchanged. For linearly polarized scattering, the vectorial singularities correspond to directions where the scattering is zero with integer Poincar\'{e} indices; while for elliptically polarized scattering, line singularities with half-integer Hopf indices correspond to directions of either zero scattering or circularly polarized scattering (denoted as \textbf{C} points). For both types of singularities, the long axis is not well defined~\cite{NYE_natural_1999,GBUR_2016__Singular}. Apart from those differences, the Poincar\'{e}-Hopf theorem is applicable to both vector fields and line fields~\cite{MILNOR_1997__Topology,NEEDHAM_1998__Visual,HOPF_2003__Differential}. Considering that the application of the theorem requires only the continuity of the scattered fields that has nothing to do with how the scatterings are induced, and that there are two types of line singularities, it is easy to conclude the following for arbitrary scattering bodies in a homogeneous background: \textit{(i) There must be isolated  singularities (singular directions on the momentum sphere) where the scattering is either zero or circularly polarized; (ii) The Hopf index sum of all those singularities has to be $2$.} From the perspective of line fields, even intrinsic vectorial singularities of Poincar\'{e} index $\pm1$ are not fundamental anymore, as they can be further separated, through adding perturbations of extra multipolar terms,  into a pair of $\textbf{C}$ points with opposite handedness and the same Hopf index of $1/2$ or $-1/2$~\cite{jackson1962classical,NYE_natural_1999,GBUR_2016__Singular}. In contrast, intrinsic line singularities of index $\pm 1/2$ are fundamental and can not possibly be further decomposed.

\begin{figure}[tp]
\centerline{\includegraphics[width=8.5cm]{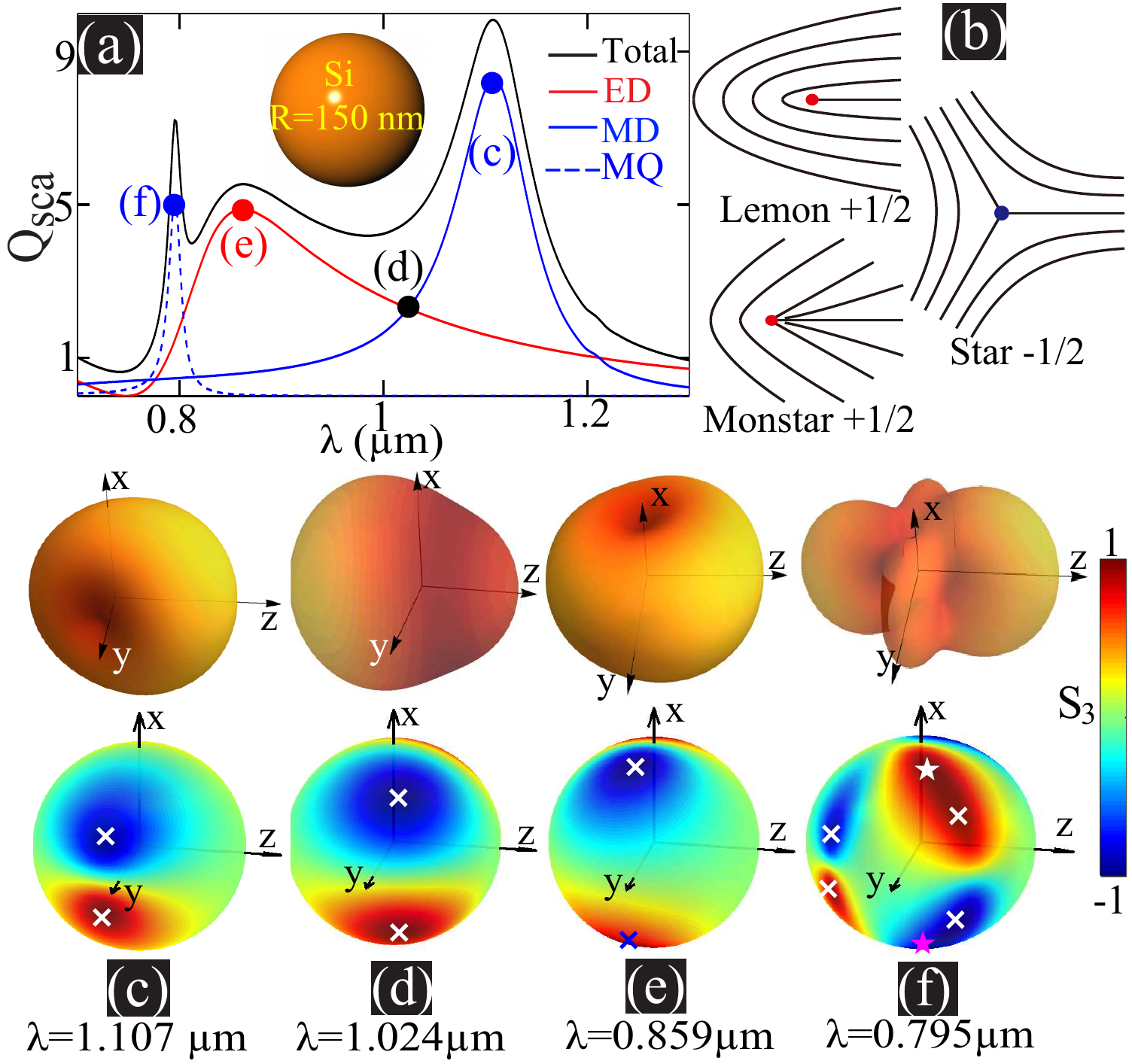}} \caption{\small Scattering efficiency spectra (both total and  partial ones contributed by the ED, MD, and MQ are shown) for a silicon sphere of radius $150$~nm. Three resonant ($\lambda=1107$~nm, $859$~nm, $795$~nm) and one non-resonant ($\lambda=1024$~nm) spectral positions are pinpointed. The scattering intensity patterns and Stokes parameter ($S_3$) distributions at those positions are shown correspondingly in (c)-(f), where \textbf{C} points are marked by crosses of Hopf index $+1/2$ or stars of Hopf index $-1/2$.  Three typical line field patterns close to intrinsic singularities are shown in (b): Lemon and Monstar singularities of Hopf index $+1/2$ and Star singularity of Hopf index $-1/2$.}
\label{fig2}
\end{figure}

To further exemplify the aforementioned basic concepts and global topological invariance of line fields constructed from general elliptically polarized scattering, we show in Fig.~\ref{fig2}(a) the scattering spectra of a silicon sphere of radius $150$~nm. Besides the contributions from the ED and MD, that from the magnetic quadrupole (MQ) is also included. At each pinpointed position, besides the scattering patterns, we show also the distributions of Stokes parameter $S_3$ on the momentum sphere to identify the positions of \textbf{C}-point singularities ($S_3=\pm 1$ corresponds respectively to left-handed and right-handed circular polarizations; $S_3=0$ corresponds to linear polarizations, as is the case throughout the momentum sphere in Fig.~\ref{fig1}) in Figs.~\ref{fig2}(c)-(f).

For the primitive Mie scattering with linearly polarized plane waves and spherical particles, linearly polarized scattering all across the momentum sphere can be obtained when Mie scattering coefficients ($a_m$ and $b_m$, where $m$ is a positive integer) fulfill one of the following requirements~\cite{Bohren1983_book}: (i) Only one coefficient is nonzero; (ii) The ratios between any two nonzero scattering coefficients are real. For the well known two cases of ED and MD scattering discussed already, the first requirement is satisfied with $a_1\neq0$ and $b_1\neq0$, respectively; while for the scenario of Kerker dipole mentioned above, the second condition is met with $a_1/b_1=1$.  Though in Fig.~\ref{fig1} we have shown the results only at the resonant spectral positions, this criterion is generally applicable. For example, Rayleigh scattering occurs at the regime where $a_1\neq0$ and the scattering is far from the resonant position ($|a_1| \ll 1$). As a result, Rayleigh scattering with linear polarized plane wave has a typical linear polarization pattern of an ED.

For the marked points shown in Fig.~\ref{fig2}(a) [including both resonant ($\lambda=1107$~nm, $859$~nm, $795$~nm) and non-resonant ($\lambda=1024$~nm) spectral positions], neither of the above aforementioned two conditions is satisfied. This can be simply confirmed through direct calculations for  ratios of the non-vanishing Mie scattering coefficients. The corresponding scattering patterns at those points render an alternative confirmation: there are no directions along which the scattering is zero [see Figs.~\ref{fig2}(c)-(f)]. This is because, as we have argued above, if the scattering is everywhere on the momentum sphere linearly polarized, there must be directions of zero scattering (\textbf{V} points). As a result, the singularities must be $\textbf{C}$ points along which the scattering is circularly polarized. Those line singularities are marked by crosses of Hopf index $+1/2$ or stars of Hopf index $-1/2$  on the $S_3$ distribution graphs shown in Figs.~\ref{fig2}(c)-(f). At each position, only half of singularities are visible and the other half can be inferred from the symmetry of the scattering with respect to the \textbf{x}-\textbf{z} plane.  We emphasize that the pseudo-scalar nature of chirality~\cite{jackson1962classical,BIRSS_1964} means that each \textbf{C} point marked in Figs.~\ref{fig2}(c)-(f) has another unshown (on the other side of the sphere) imaging singularity partner with opposite handedness ($S_3=\pm1$) while the same Hopf index (mirror symmetry does not change the Hopf index of the singularity). The common feature of the three dipolar (both resonant and non-resonant) scattering cases [shown in Figs.~\ref{fig2}(c)-(e), where there are no quadrupolar or higher-order multipolar contributions] is that, there are four \textbf{C} points across the momentum sphere and each has a Hopf index of $+1/2$  with the sum being $2$. For the last case with dominant MQ scattering ($\lambda=795$~nm), on the momentum sphere there are altogether twelve isolated singularities (only six are visible): four of Hopf index $-1/2$ (two of them are marked by stars) and eight of Hopf index $+1/2$ (four of them are marked by crosses), with the index sum being $2$.

It is worth mentioning that, for both vectorial and line singularities, each index corresponds to infinitely many possible field patterns that are geometrically distinct while topologically equivalent at the neighbourhood of the singularities~\cite{NYE_natural_1999,GBUR_2016__Singular,GALVEZ_2014_Phys.Rev.A_Generation}. In Fig.~\ref{fig2}(b) we show only three typical line field patterns close to the intrinsic singularities of Hopf index of $\pm 1/2$, that is, Lemon and Monstar singularities of Hopf index $+1/2$ and Star singularity of Hopf index $-1/2$. The Lemon and Monstar singularities have the same Hopf index and thus topologically equivalent, meaning that they are smoothly interconvertible~\cite{GALVEZ_2014_Phys.Rev.A_Generation,KUMAR_2015_ComplexLightOpt.ForcesIX_Monstar}. We further note that investigations into the scattering properties of dielectric particles from the perspective of line field and line singularities have been conducted before~\cite{GARCIA-ETXARRI_2017_ACSPhotonics_Opticala}, which nevertheless concentrate on the local properties, without revealing the global feature of unavoidable existences of singularities (isolated directions of zero or circularly polarized scattering) and the invariance of the index sum. This is understandable, since previous studies were not performed within the general global framework of the Poincar\'{e}-Hopf theorem.

\begin{figure}[tp]
\centerline{\includegraphics[width=8.5cm]{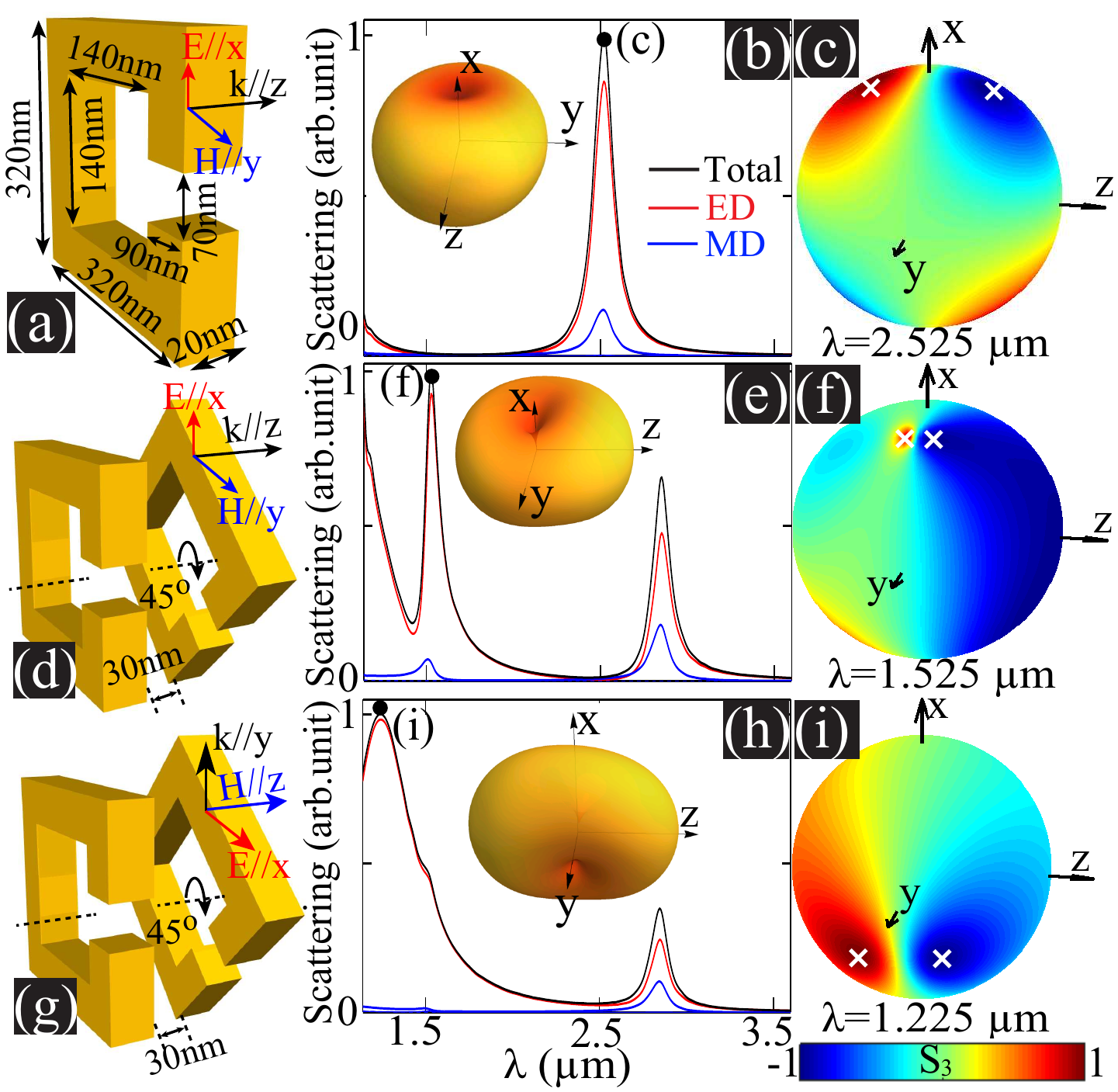}} \caption{\small Linear polarized plane waves scattered by an indiviudal SRR in (a), or twisted SRR pairs in (d) and (g). All SRRs are identical with the geometric parameters specified in (a).  The twisted pair consist of two \textbf{z}-coaxial SRRs with a $45^\circ$ relative rotation and a $30$~nm gap between them. The scattering spectra (normalized) are shown correspondingly in (b), (e) and (h). For each scattering configuration, a resonant position is marked ($\lambda=2.525~\mu$m, $1.525~\mu$m and $1.225~\mu$m), at which the scattering intensity patterns are included as insets. The $S_3$ distributions are shown respectively in (c), (f) and (i), and for all of them there are four \textbf{C}-point singularities of Hopf index $+1/2$ (half of them are visible and marked by crosses), with the index sum being $2$.}\label{fig3}
\end{figure}

\section{Arbitrarily Polarized Scattering by Particle Clusters}

As mentioned already, our analysis based on Poincar\'{e}-Hopf theorem has nothing to do with the shape of the scattering bodies (being them ideally spherical or not), or how those scattering bodies are clustered and interact with one another. As a next step, we turn to the geometrically less symmetric while widely celebrated structure of split ring resonators (SRRs) for further illustrations of the global Mie scattering.   The SRRs investigated in this work are identical, geometric and optical parameters of which are the same as those in the experimental study: geometric parameters are specified in Fig.~\ref{fig3}(a); they are made of gold, the permittivity of which in our investigated spectral regime is given by Drude model ${\varepsilon}(\omega ) = 1 -
\omega_p^2/\omega({\omega } + i {\omega _c })$, where ${\omega_p}\approx 1.37\times10^{16}$~Hz is the plasma frequency and ${\omega _c}\approx 4.08\times10^{13}$~Hz is the collision frequency~\cite{Linden2004_science}. We firstly study the widely employed configurations with linearly polarized incident plane waves shown in Figs.~\ref{fig3}(a), (d) and (g), including both an individual SRR and a twisted SRR pair [two \textbf{z}-axis-coaxial SRRs with a $45^\circ$ relative rotation and  a $30$~nm gap, as shown in Figs.~\ref{fig3}(d) and (g)].  The spectral regime examined is where there is significant optically-induced magnetic response that is foundational for the field of metamaterials, with the corresponding scattering spectra of each configuration shown correspondingly in Figs.~\ref{fig3}(b), (e) and (h).  For each case we have marked a resonant position ($\lambda=2.525~\mu$m, $1.525~\mu$m and $1.225~\mu$m, respectively), at which the $S_3$ parameter distributions across the momentum sphere are shown respectively in Figs.~\ref{fig3}(c), (f) and (i). Similar to those shown in Fig.~\ref{fig2}, none scenario manifests \textbf{V}-point singularities along which there are no radiations [see scattering intensity patterns included as insets of Figs.~\ref{fig3}(b), (e) and (h)]. This is natural, as we have already argued above, \textbf{V}-point singularities are accidental and would easily break up into  intrinsic \textbf{C}-point singularities for general elliptically polarized scatterings. For each case, four intrinsic \textbf{C}-point singularities of Hopf index $+1/2$ (half of them are visible and pinpointed by crosses) are always present [see Figs.~\ref{fig3}(c), (f) and (i)], with the index sum being always $2$, as dictated by the Poincar\'{e}-Hopf theorem.

\begin{figure}[tp]
\centerline{\includegraphics[width=8.2cm]{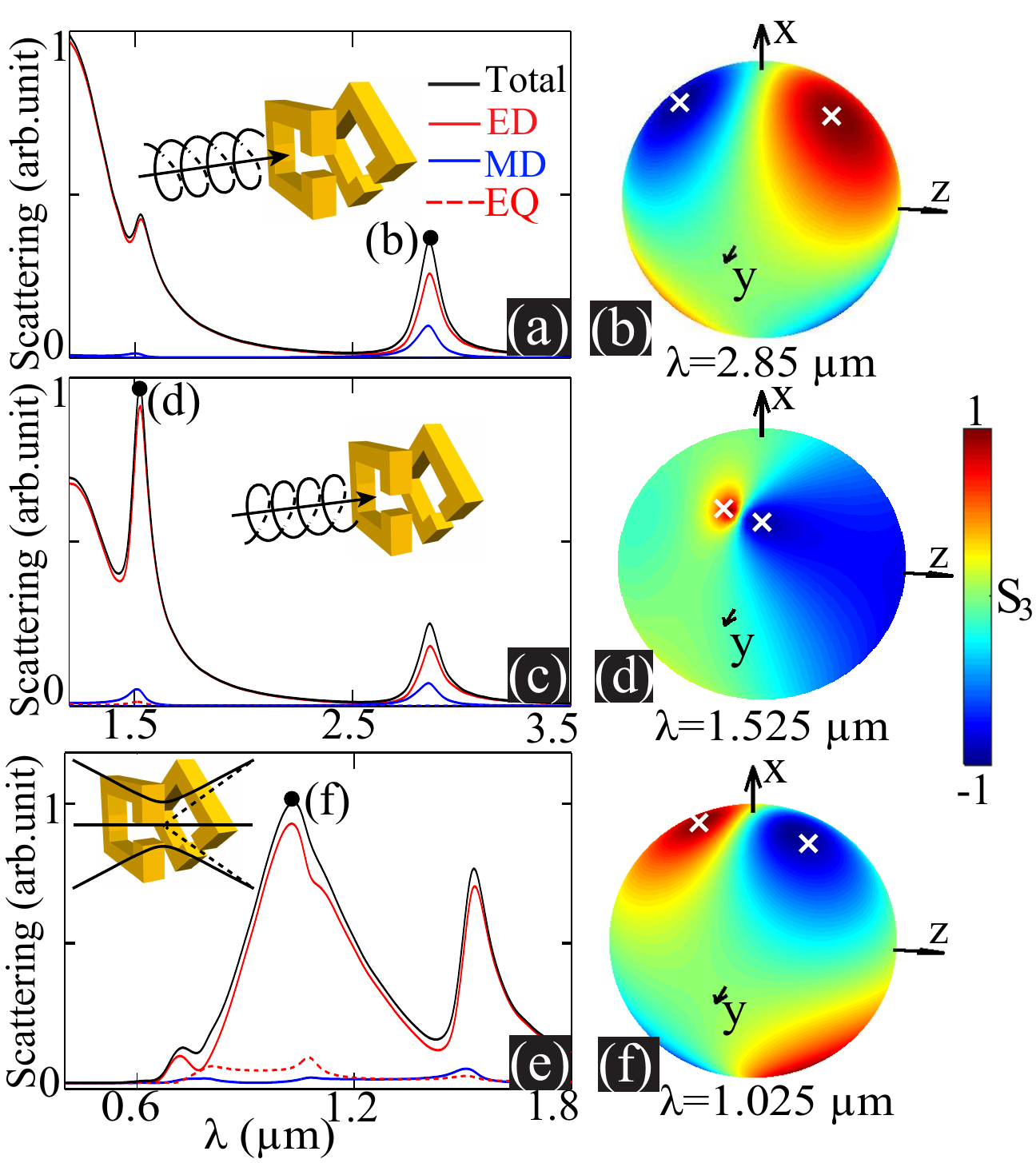}} \caption{\small The scattering of left and right handed circularly poalrized planes waves, and a Gaussian beam by the SRR pairs that are the same (see insets) as those studied already in Fig.~\ref{fig3}, with the spectra shown respectively in (a), (c) and (e). A resonant position is marked for each case ($\lambda=2.85~\mu$m, $1.525~\mu$m and $1.025~\mu$m), at which the $S_3$ distributions are shown respectively in (b), (d) and (f). For all scenarios there are four \textbf{C}-point singularities of Hopf index $+1/2$ (half of them are visible and marked by crosses), with the index sum being $2$.}
\label{fig4}
\end{figure}

Though we have so far confined our discussions to the incidence of linearly polarized plane waves, our conclusion concerning the global topological properties are independent on not only scattering bodies, but also the incident waves. Now we consider the same configuration of twisted SRR pair [as is shown in Fig.~\ref{fig3}(d)] that is widely employed to study artificial optical activity~\cite{Soukoulis2011_NP,HENTSCHEL_2017_Sci.Adv._Chiral}, but switch the incident waves to circular polarizations.  The scattering spectra shown in  Figs.~\ref{fig4}(a) and (c) for incident waves of different handedness (left and right handed, respectively) are distinct due to the chirality of the SRR pair. For each case we have marked a resonant position ($\lambda=2.85~\mu$m and $1.525~\mu$m) at which the $S_3$ parameter distributions across the momentum are shown accordingly in Figs.~\ref{fig4}(b) and (d). As is clearly demonstrated, despite the opposite handedness, for both cases there are four intrinsic \textbf{C}-point singularities of Hopf index $+1/2$ (half of them are visible and marked by crosses).  We further investigate the scattering by such a SRR pair of a x-polarized Gaussian beam, which propagates along \textbf{z} axis  with the beam waist radius $1.5~\mu$m and the beam center coinciding with the geometric center of the SRR pair. Its scattering spectra are shown in Fig.~\ref{fig4}(e) with the presence of also an electric quadrupole (EQ). At the marked resonant position ($\lambda=1.025~\mu$m), the corresponding $S_3$ parameter distribution in Fig.~\ref{fig4}(f) clearly manifests \textbf{C}-point singularities (two of four singularities are visible and marked; all with Hopf index $+1/2$), indicating that the global topological properties are preserved regardless of the spatial shape of the incident wave. We emphasize that here in Figs.~\ref{fig3} and \ref{fig4} we have investigated only the long wavelength regime when there are only dominant lower-order multipolar (up to quadrupole) scattering. As for the larger frequency regions with the emergence of higher order multipoles, more singularities will be observed across the momentum sphere [harder to be directly visualized though; see for example Fig.~\ref{fig2}(f)]. The index sum of all singularities are bounded to be $2$, no matter how many multipoles are involved and how high the orders of the involved multipoles can be. This principle holds valid at any spectral positions including non-resonant ones [see for example Fig.~\ref{fig2}(d)], although in Figs.~\ref{fig3} and \ref{fig4} we have shown only the results at the resonant points where the scattering is stronger and easier to be measured.

\section{Conclusions and Outlook}

Our work here has addressed a fundamental problem: For light scattering by arbitrary obstacles, is it possible to define globally invariant properties to characterize any scattering that is dependent on neither the scattering bodies nor the incident waves? Through a cornerstone theorem of global differential geometry, the Poincar\'{e}-Hopf theorem, we show that for arbitrary finite obstacle scatterings in a homogenous background, there must be isolated singular directions where the scattering is either zero or circularly polarized. For both sorts of singularities, we can assign half-integer Hopf indices and the index sum of all singularities has to be $2$, which is the Euler characteristic of the momentum sphere. Since the Poincar\'{e}-Hopf theorem requires only the continuity of the tangent fields (that is transverse electromagnetic fields scattered to the far field for our specific problem), it cares nothing about how the fields are generated and thus the global properties we have revealed are generically independent of the specific scattering configurations. Here for the first time we have merged Mie scattering with global topological concepts, which can potentially nourish new perspectives for both conventional Mie scattering related problems, such as optical resonators, metasurfaces, vectorial vortex beams and so on,  but also the recently rapidly expanding field of topological photonics. Since scattering is a fundamental phenomena for different branches of physics, which pervades not only electromagnetic waves but also waves of other natures,  we believe our work also shed new light on subjects such as acoustics, seismic studies and other subdisciplines involving matter waves.

At the same time, we would like to point out the limitations of our work: (i) Here we have confined our discussions to far-field scatterings, where the fields are transverse (tangent) and thus the Poincar\'{e}-Hopf theorem is directly applicable. In the near field however, the electromagnetic fields are not transverse anymore and the employment of the Poincar\'{e}-Hopf theorem requires that all radial terms of the fields are dropped, and then the conclusions we have drawn are still valid for arbitrary near-field scattered or radiated electromagnetic waves; (ii) Neither static vector fields nor line fields can be directly defined for partially polarized and/or incoherent light, and then the global principle we have discovered here is not directly applicable;  (iii) Singularities in terms of either electric or magnetic fields (as we have done here) are not invariant under Lorentz transformations, and thus for more general scenarios (such as scattering within a moving background; or the scattering body and the observer are in relative motion) the introductions of Riemann-Silberstein fields are necessary~\cite{BIALYNICKI-BIRULA_2003_Phys.Rev.A_Vortex,BERRY_2004_J.Opt.A:PureAppl.Opt._Riemann}. At the current stage, it is not clear whether or not it is possible to define invariant parameters to characterize scattering of those more general scenarios, which deserves further investigations from such a global perspective.

\section*{acknowledgement}
We acknowledge the financial support from National Natural Science
Foundation of China (Grant No. 11874026, 11404403 and 11874426), and the Outstanding Young Researcher Scheme of National University of Defense Technology.

W. Chen and Q. Yang contributed equally to this work.

\section*{Appendix: Methods}

For spherical particles (both homogeneous and core-shell ones), the scattering spectra (both total scattering and partial scattering from different multipolar components), angular radiation patterns and Stokes parameter ($S_3$) distributions can be analytically calculated from standard Mie theory~\cite{Bohren1983_book}. With the scattered field distributions obtained close to the singularities, the indices (both Poincar\'{e} and Hopf ones) can be directly calculated through the approaches described in Ref.~\cite{BERRY_2004_J.Opt.PureAppl.Opt._Index}.

For non-spherical structures studied in this work, we employ a commercial software package COMSOL MULTIPHYSICS (https://www.comsol.com) to extract the scattered fields. With the extracted information we can directly calculate the total scattering spectra, angular radiation patterns and Stokes parameter distributions. For partial scattering spectra of different multipolar components, we conduct the spherical harmonic expansions of the extracted scattered fields to obtain all the expansion coefficients:  $a_{nm}$ for electric multipoles and $b_{nm}$ for magnetic multipoles~\cite{GRAHN_NewJ.Phys._electromagnetic_2012}. With those coefficients, the total scattering can be accurately separated and attributed to different multipoles~\cite{CHEN_2019_LaserPhotonicsRev._Multipolara}: the scattering contributed by electric multipoles of order $n$ is proportional to ${\sum\nolimits_{m=-n}^{m=n} (2n+1){|{a_{\rm{nm}}}|} ^2}$ and that contributed by magnetic multipoles is proportional to ${\sum\nolimits_{m=-n}^{m=n} (2n+1) {|{b_{\rm{nm}}}|} ^2}$. For example, the scatting of ED and MD is proportional to ${\sum\nolimits_{m=-1}^{m=1} 3{|{a_{\rm{1m}}}|} ^2}$ and ${\sum\nolimits_{m=-1}^{m=1} 3{|{b_{\rm{1m}}}|} ^2}$, respectively; while the scattering of EQ and MQ is proportional to ${\sum\nolimits_{m=-2}^{m=2} 5{|{a_{\rm{2m}}}|} ^2}$ and ${\sum\nolimits_{m=-2}^{m=2} 5{|{b_{\rm{2m}}}|} ^2}$, respectively.  We note that for the scenario of an incident Gaussian beam, we have subtracted the incident wave from the total extracted fields in COMSOL to calculate all the scattering parameters.


\begin{thebibliography}{10}
\newcommand{\enquote}[1]{``#1''}

\bibitem{Bohren1983_book}
C.~F. Bohren and D.~R. Huffman, \emph{Absorption and Scattering of Light by
  Small Particles} (Wiley, 1983).

\bibitem{DOICU_light_2006}
A.~Doicu, T.~Wriedt, and Y.~A. Eremin, \emph{Light Scattering by Systems of
  Particles: Null-Field Method with Discrete Sources: Theory and Programs},
  vol. 124 ({Springer}, 2006).

\bibitem{GOUESBET_generalized_2011}
G.~Gouesbet and G.~Gr{\'e}han, \emph{Generalized {{Lorenz}}-{{Mie Theories}}}
  ({Springer Science \& Business Media}, 2011).

\bibitem{FERRARI_2005_NatRevCancer_Cancer}
M.~Ferrari, \enquote{Cancer nanotechnology: Opportunities and challenges,} Nat.
  Rev. Cancer \textbf{5}, 161--171 (2005).

\bibitem{DREADEN_2012_Chem.Soc.Rev._golden}
E.~C. Dreaden, A.~M. Alkilany, X.~Huang, C.~J. Murphy, and M.~A. {El-Sayed},
  \enquote{The golden age: Gold nanoparticles for biomedicine,} Chem. Soc. Rev.
  \textbf{41}, 2740--2779 (2012).

\bibitem{Alu2009_PRL}
A.~Alu and N.~Engheta, \enquote{Cloaking a sensor,} Phys. Rev. Lett.
  \textbf{102}, 233901 (2009).

\bibitem{Ruan2010_PRL}
Z.~C. Ruan and S.~H. Fan, \enquote{Superscattering of light from subwavelength
  nanostructures,} Phys. Rev. Lett. \textbf{105}, 013901 (2010).

\bibitem{QIAN_2019_Phys.Rev.Lett._Experimental}
C.~Qian, X.~Lin, Y.~Yang, X.~Xiong, H.~Wang, E.~Li, I.~Kaminer, B.~Zhang, and
  H.~Chen, \enquote{Experimental {{Observation}} of {{Superscattering}},} Phys.
  Rev. Lett. \textbf{122}, 063901 (2019).

\bibitem{Kerker1983_JOSA}
M.~Kerker, D.~S. Wang, and C.~L. Giles, \enquote{Electromagnetic scattering by
  magnetic spheres,} J. Opt. Soc. Am. \textbf{73}, 765 (1983).

\bibitem{LIU_2018_Opt.Express_Generalized}
W.~Liu and Y.~S. Kivshar, \enquote{Generalized {{Kerker}} effects in
  nanophotonics and meta-optics {{[Invited]}},} Opt. Express \textbf{26},
  13085--13105 (2018).

\bibitem{jahani_alldielectric_2016}
S.~Jahani and Z.~Jacob, \enquote{All-dielectric metamaterials,} Nat.
  Nanotechnol. \textbf{11}, 23--26 (2016).

\bibitem{KUZNETSOV_Science_optically_2016}
A.~I. Kuznetsov, A.~E. Miroshnichenko, M.~L. Brongersma, Y.~S. Kivshar, and
  B.~Luk'yanchuk, \enquote{Optically resonant dielectric nanostructures,}
  Science \textbf{354}, aag2472 (2016).

\bibitem{WON_2019__Mietronic}
R.~Won, \enquote{Into the `{{Mie}}-tronic' era,} Nat. Photonics \textbf{13},
  585--587 (2019).

\bibitem{Lu2014_topological}
L.~Lu, J.~D. Joannopoulos, and M.~Soljacic, \enquote{Topological photonics,}
  Nat. Photonics \textbf{8}, 821 (2014).

\bibitem{OZAWA_2018_ArXiv180204173}
T.~Ozawa, H.~M. Price, A.~Amo, N.~Goldman, M.~Hafezi, L.~Lu, M.~C. Rechtsman,
  D.~Schuster, J.~Simon, O.~Zilberberg, and I.~Carusotto, \enquote{Topological
  photonics,} Rev. Mod. Phys. \textbf{91}, 015006 (2019).

\bibitem{CHEN_2019__Singularities}
W.~Chen, Y.~Chen, and W.~Liu, \enquote{Singularities and poincar\'e indices of
  electromagnetic multipoles,} Phys. Rev. Lett. \textbf{122}, 153907 (2019).

\bibitem{CHEN_2019_ArXiv190409910Math-PhPhysicsphysics_Linea}
W.~Chen, Y.~Chen, and W.~Liu, \enquote{Line {{Singularities}} and {{Hopf
  Indices}} of {{Electromagnetic Multipoles}},} arXiv p. 1904.09910 (2019).

\bibitem{MILNOR_1997__Topology}
J.~W. Milnor, \emph{Topology from the {{Differentiable Viewpoint}}} ({Princeton
  University Press}, Princeton, N.J, 1997).

\bibitem{NEEDHAM_1998__Visual}
T.~Needham, \emph{Visual {{Complex Analysis}}} ({Clarendon Press}, 1998).

\bibitem{Johnson1972_PRB}
P.~B. Johnson and R.~W. Christy, \enquote{Optical constants of the noble
  metals,} Phys. Rev. B \textbf{6}, 4370 (1972).

\bibitem{Maier2007}
S.~A. Maier, \emph{Plasmonics : fundamentals and applications} (Springer, New
  York, 2007).

\bibitem{jackson1962classical}
J.~D. Jackson, \emph{Classical Electrodynamics} (Wiley New York, 1962).

\bibitem{PALIK_1998__Handbook}
E.~D. Palik, \emph{Handbook of Optical Constants of Solids}, vol.~3 ({Academic
  press}, 1998).

\bibitem{liu_toroidal_2015}
W.~Liu, J.~Zhang, and A.~E. Miroshnichenko, \enquote{Toroidal dipole-induced
  transparency in core\textendash{}shell nanoparticles,} Laser Photonics Rev.
  \textbf{9}, 564--570 (2015).

\bibitem{FENG_2017_Phys.Rev.Lett._Ideal}
T.~Feng, Y.~Xu, W.~Zhang, and A.~E. Miroshnichenko, \enquote{Ideal {{Magnetic
  Dipole Scattering}},} Phys. Rev. Lett. \textbf{118}, 173901 (2017).

\bibitem{Liu2012_ACSNANO}
W.~Liu, A.~E. Miroshnichenko, D.~N. Neshev, and Y.~S. Kivshar,
  \enquote{Broadband unidirectional scattering by magneto-electric core-shell
  nanoparticles,} ACS Nano \textbf{6}, 5489 (2012).

\bibitem{HOPF_2003__Differential}
H.~Hopf, \emph{Differential {{Geometry}} in the {{Large}}: {{Seminar Lectures
  New York University}} 1946 and {{Stanford University}} 1956} ({Springer},
  2003).

\bibitem{NYE_natural_1999}
J.~F. Nye, \emph{Natural {{Focusing}} and {{Fine Structure}} of {{Light}}:
  {{Caustics}} and {{Wave Dislocations}}} ({CRC Press}, 1999).

\bibitem{GBUR_2016__Singular}
G.~J. Gbur, \emph{Singular {{Optics}}} ({CRC Press Inc}, Boca Raton, 2016).

\bibitem{BIRSS_1964}
R.~R. Birss, \emph{Symmetry and {{Magnetism}}} ({North-Holland Publishng},
  1964), 1st ed.

\bibitem{GALVEZ_2014_Phys.Rev.A_Generation}
E.~J. Galvez, B.~L. Rojec, V.~Kumar, and N.~K. Viswanathan, \enquote{Generation
  of isolated asymmetric umbilics in light's polarization,} Phys. Rev. A
  \textbf{89}, 031801(R) (2014).

\bibitem{KUMAR_2015_ComplexLightOpt.ForcesIX_Monstar}
V.~Kumar and N.~K. Viswanathan, \enquote{Is {{Monstar}} topologically the same
  as lemon?} in \enquote{Complex {{Light}} and {{Optical Forces IX}},} , vol.
  9379 ({International Society for Optics and Photonics}, 2015), vol. 9379, p.
  937909.

\bibitem{GARCIA-ETXARRI_2017_ACSPhotonics_Opticala}
A.~{Garcia-Etxarri}, \enquote{Optical polarization mobius strips on
  all-dielectric optical scatterers,} ACS Photonics \textbf{4}, 1159--1164
  (2017).

\bibitem{Linden2004_science}
S.~Linden, C.~Enkrich, M.~Wegener, J.~F. Zhou, T.~Koschny, and C.~M. Soukoulis,
  \enquote{Magnetic response of metamaterials at 100 terahertz,} Science
  \textbf{306}, 1351 (2004).

\bibitem{Soukoulis2011_NP}
C.~M. Soukoulis and M.~Wegener, \enquote{Past achievements and future
  challenges in the development of three-dimensional photonic metamaterials,}
  Nat. Photonics \textbf{5}, 523--530 (2011).

\bibitem{HENTSCHEL_2017_Sci.Adv._Chiral}
M.~Hentschel, M.~Sch{\"a}ferling, X.~Duan, H.~Giessen, and N.~Liu,
  \enquote{Chiral plasmonics,} Sci. Adv. \textbf{3}, e1602735 (2017).

\bibitem{BIALYNICKI-BIRULA_2003_Phys.Rev.A_Vortex}
I.~{Bialynicki-Birula} and Z.~{Bialynicka-Birula}, \enquote{Vortex lines of the
  electromagnetic field,} Phys. Rev. A \textbf{67}, 062114 (2003).

\bibitem{BERRY_2004_J.Opt.A:PureAppl.Opt._Riemann}
M.~V. Berry, \enquote{Riemann\textendash{{Silberstein}} vortices for paraxial
  waves,} J. Opt. A: Pure Appl. Opt. \textbf{6}, S175--S177 (2004).

\bibitem{BERRY_2004_J.Opt.PureAppl.Opt._Index}
M.~V. Berry, \enquote{Index formulae for singular lines of polarization,} J.
  Opt. A: Pure Appl. Opt. \textbf{6}, 675 (2004).

\bibitem{GRAHN_NewJ.Phys._electromagnetic_2012}
P.~Grahn, A.~Shevchenko, and M.~Kaivola, \enquote{Electromagnetic multipole
  theory for optical nanomaterials,} New J. Phys. \textbf{14}, 093033 (2012).

\bibitem{CHEN_2019_LaserPhotonicsRev._Multipolara}
W.~Chen, Y.~Chen, and W.~Liu, \enquote{Multipolar {{Conversion Induced
  Subwavelength High}}-{{Q Kerker Supermodes}} with {{Unidirectional
  Radiations}},} Laser Photonics Rev. \textbf{13}, 1900067 (2019).

\end{thebibliography}

\end{document}